\documentclass[aps,pre,twocolumn,showpacs,superscriptaddress,preprintnumbers,amsmath,amssymb]{revtex4}
\usepackage{amsmath}
\usepackage{graphicx}
\usepackage{dcolumn}
\usepackage{bm}
\usepackage{float}

\begin{document}
\preprint{\textit{Preprint}}
\title{Radiation from laser-microplasma-waveguide interactions in the ultra-intense regime}
\author{Longqing Yi}\thanks{Authors to whom correspondence should be addressed. Electronic mail: yi@uni-duesseldorf.de}
\affiliation{Institut f$\ddot{u}$r Theoretische Physik I, Heinrich-Heine-Universit$\ddot{a}$t D$\ddot{u}$sseldorf, D$\ddot{u}$sseldorf, 40225 Germany}\affiliation{State Key Laboratory of High Field Laser Physics, Shanghai Institute of Optics and Fine Mechanics, Chinese Academy of Sciences, P.O. Box 800-211, Shanghai 201800, China}
\author{Alexander Pukhov}
\affiliation{Institut f$\ddot{u}$r Theoretische Physik I, Heinrich-Heine-Universit$\ddot{a}$t D$\ddot{u}$sseldorf, D$\ddot{u}$sseldorf, 40225 Germany}
\author{Baifei Shen}
\affiliation{State Key Laboratory of High Field Laser Physics, Shanghai Institute of Optics and Fine Mechanics, Chinese Academy of Sciences, P.O. Box 800-211, Shanghai 201800, China}
\affiliation{Collaborative Innovation Center of IFSA (CICIFSA),Shanghai Jiao Tong University, Shanghai 200240, China}
\date{\today}
\newcommand{\mum}{~$\mu$m}
\newcommand{\wcm}{~W/cm$^{2}$}
\newcommand{\gcm}{~g/cm$^{3}$}
\newcommand{\ccm}{~cm$^{-3}$}
\newcommand{\TVM}{~TV/m}
\newcommand{\Carbon}{C$^{6+}$}
\newcommand{\Gold}{Au$^{1+}$}
\newcommand{\Carbontwo}{C$^{2+}$}
\newcommand{\Hydrogen}{H$^{+}$}

\setlength{\belowdisplayskip}{5pt} \setlength{\belowdisplayshortskip}{0pt}
\setlength{\abovedisplayskip}{5pt} \setlength{\abovedisplayshortskip}{0pt}

\begin{abstract}
When a high-contrast ultra-relativistic laser beam enters a micro-sized plasma waveguide, the pulse energy is coupled into waveguide modes, which remarkably modifies the interaction of electrons and electromagnetic wave. The electrons that pulled out of walls form a dense helical bunch inside the channel are efficiently accelerated by the transverse magnetic modes to hundreds of MeVs. In the mean time, the asymmetry in the transverse electric and magnetic fields provides significant wiggling that leads to a bright, well-collimated emission of hard X-rays. In this paper, we present our study on the underlying physics in the aforementioned process using 3D particle-in-cell simulations. The mechanism of electron acceleration and the dependence of radiation properties on different laser plasma parameters are addressed. A theoretical analysis model and basic scalings for X-ray emission are also presented by considering the lowest optical modes in the waveguide, which is adequate to describe the basic observed phenomenon. In addition, the effects of high order modes as well as laser polarization are also qualitatively discussed. The considered X-ray source have promising features that might serve as a competitive candidate for future tabletop synchrotron source.
\end{abstract}
\maketitle

\section*{I. INTRODUCTION}

The interaction of ultra-intense laser pulses with matter is entering a new era owing to the advances in laser pulse cleaning techniques \cite{s1,s2} and manufacturing structured targets in micro- and nano-scale \cite{s3,s4,s5}. Using cross-polarized wave generation technique, laser peak-to-pedestal contrast ratios higher than 10$^{10}$ have been achieved \cite{s1}, which allows for interaction with fine plasma structures without destroying the guiding features before the arrival of the intense portion of the pulse. Plasma structures such as gratings \cite{s6}, nanograsses \cite{s7}, snowflakes \cite{s8}, and channels \cite{s9} have been studied with the aim of enhancing laser absorption and manipulating laser-matter interaction. It has been reported experimentally that significant modifications to many physical processes, including the acceleration of particles \cite{s4,s9} and secondary radiations \cite{s3,s7}, are induced by these guiding structures. Nevertheless, simulations have also suggested many interesting phenomenons \cite{s5,s10,s11,s12,s13} can be achieved with the development of laser technology in the near future that may lead to diverse potential applications in fundamental science, industry, and medicine.

Among these structured targets of interest is the micro plasma waveguide (MPW) \cite{s10}, which brings promising features to the study of laser-plasma based X-ray sources. Previously, prior researches have mainly been focused on the radiation from electrons generated by laser-driven wakefield accelerators (LWFAs) \cite{s14,s15,s16} as well as hot electrons produced in interactions of high power laser and foils \cite{s17,s18,s19,s20,s21} or near-critical-density (NCD) plasmas \cite{s22,s23,s24}. However, although the x-ray source based on LWFA is more mature, the electron number produced by the LWFA is limited by the underdense plasma required for wakefield excitation, which further results in a relatively-small photon yield. On the other hand, the number of hot electron generated in laser-foil or laser-NCD-plasma interaction is much more larger and a much higher conversion efficiency can be expected. But the typical divergence for X-rays produced in this case is several tens of degrees, making it unsuitable for many applications.

On the other hand, simulations has suggests the X-ray source based on a MPW presents promising features in both aspects \cite{s10}. In this case, overcritical electron bunches can be obtained \cite{s25}. The large number of electrons extract significant laser energy during the co-propagation with laser pulse for hundreds of micrometers, and convert it into abundant X-ray photon emission. The asymmetry in the transverse electric and magnetic fields is an essential aspect of the optical modes inside a MPW, by adjusting the laser plasma parameters, one can make sure the transverse force is sufficiently large to produce high harmonics that reaches X-ray frequency regime and, at the same time, small enough to maintain a good divergence.

Although laser-waveguide interaction below material damage threshold is well understood \cite{s26,s27}, and many prior researches has been done for laser propagating in plasma waveguides \cite{s28,s29,s30}, the physics of relativistic intense laser pulses interaction with electrons inside a MPW as well as the subsequent radiation process is not extensively explored. Therefore, it is necessary to develop predictive models for radiation properties based on theoretical analysis and particle-in-cell (PIC) simulations. In this paper, we perform full (3D) PIC simulations that studies the radiation spectrum, angular distribution, X-ray photon yield as well as brilliance of radiation in detail. Parameter dependence and several practical issues such as laser polarization and the excitation of high order modes are considered in order to guide the forthcoming experiments and future tabletop X-ray source development that based on laser-plasma-waveguide interactions.

This paper is organized as follows: Sec.~II gives a brief description on the theoretical model we employed to analyze the optical modes inside a MPW; the acceleration process of electron bunches that pulled into the MPW are discussed in Sec.~III; The radiation properties of X-rays as well as the dependence on laser-plasma parameters and the effects induced by the high order modes and laser polarization are addressed in Sec.~IV; finally, a brief conclusion is given in Sec.~V.

\section*{II. THEORETICAL MODEL}

The sketch of X-ray source based on ultra-relativistic laser pulse interacting with MPW is presented in Fig.~1 (a). A circularly polarized (CP) high power laser pulse is focused in a MPW. Electrons are extracted from the inner boundary of the MPW and accelerated by the longitudinal force due to the coupling of the laser in the waveguide modes. The resulting electron bunch displays a helical geometry due to the configuration of the accelerating field gradient. At the same time, electrons are wiggled by the transverse component of the force, producing the electromagnetic (EM) emission. Being the electrons and the laser pulse co-propagating, the wiggling occurs since the phase velocity $v_{p}$ of all the waveguide modes is superluminal. The resulting X-ray beam has a wide spectrum extending to a couple of hundreds keV and possesses a high degree of collimation.

The basic concept is making use of the special electric and magnetic field structure in the MPW as illustrated in Fig.~1(b), when a plane EM wave enters the MPW, part of the electric (or magnetic) field is coupled into the longitudinal direction to form the transverse magnetic (TM) [or transverse electric (TE)] modes. As a consequence, longitudinal electric field arises so that co-propagating electrons located at the right phase can be accelerate to relativistic energies. On the other hand, the asymmetry in transverse fields (i.e. $|E_{\bot}-B_{\bot}|\neq0$) exerts a net wiggling force on these energetic electrons, which results in synchrotron-like radiation.

\begin{figure}[!t]
\vspace{-10pt}
\includegraphics[width=8.5cm]{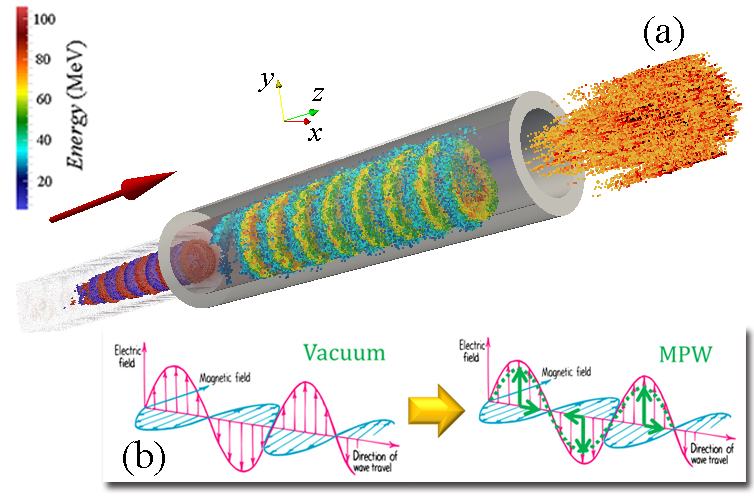}\caption{\label{f1} (Color online.) (a) Sketch map of the X-ray source based on MPW, A high power laser pulse (red-blue sequence) is focused on the entrance of MPW (white cylinder), the red arrow shows the propagation direction; a helical energetic electron bunch is generated inside the MPW co-propagating with the laser pulse, where the color of the electrons denotes the kinetic energy; these electrons are wiggled due to the asymmetry in electric and magnetic fields of the optical modes, resulting the generation of synchrotron X-ray emission (orange block). (b) Sketch of a plane EM wave in vacuum transforming into a TM mode in the MPW.}
\vspace{-20pt}
\end{figure}

In order to describe the optical modes in the MPW and provide the context for the discussion of the X-ray radiation, we briefly review the key results of the optical modes in plasma waveguide. Specifically, we numerically solve the characteristic (eigenvalue) equation and derive the scaling laws for field amplitude and phase velocity of the waveguide modes, which governs the electron acceleration and radiation processes that will be discussed in Sec.~III and IV. For simplicity, in the following, we focus on the electron interaction with the lowest modes (i.e. the first root of eigenvalue equation). The absence of high order modes causes slight deviation with the prediction of the model especially in large channels (as will be briefly discussed in Sec.~IV.C), however those difference would be in the details of the results rather than the order of magnitudes of physical quantities of interest. Therefore, it is useful to describe the qualitative features and underlying physics of the mechanism. In cylinder coordinate system ($r$, $\phi$, $z$), the electric and magnetic components of eigenmode in MPW can be written as \cite{s10}

\begin{align}
\vspace{-10pt}
\begin{split}
E_{z}=E_{z0}J_{1}(Tr)\sin(\phi)e^{-jk_{z}z}+c.c.,
\end{split}
\vspace{-10pt}
\end{align}

\begin{align}
\vspace{-10pt}
\begin{split}
B_{z}=B_{z0}J_{1}(Tr)\cos(\phi)e^{-jk_{z}z}+c.c.,
\end{split}
\vspace{-10pt}
\end{align}

\begin{align}
\vspace{-10pt}
\begin{split}
E_{r}=j\frac{k}{T}E_{z0}[\frac{J_{1}(Tr)}{Tr}-\frac{k_{z}}{k}J_{1}^{'}(Tr)]\sin(\phi)e^{-jk_{z}z}+c.c.,
\end{split}
\vspace{-10pt}
\end{align}

\begin{align}
\vspace{-10pt}
\begin{split}
B_{r}=j\frac{k}{T}B_{z0}[\frac{J_{1}(Tr)}{Tr}-\frac{k_{z}}{k}J_{1}^{'}(Tr)]\cos(\phi)e^{-jk_{z}z}+c.c.,
\end{split}
\vspace{-10pt}
\end{align}

\begin{align}
\vspace{-10pt}
\begin{split}
E_{\phi}=j\frac{k}{T}E_{z0}[J_{1}^{'}(Tr)-\frac{k_{z}}{k}\frac{J_{1}(Tr)}{Tr}]\cos(\phi)e^{-jk_{z}z}+c.c.,
\end{split}
\vspace{-10pt}
\end{align}

\begin{align}
\vspace{-10pt}
\begin{split}
B_{\phi}=-j\frac{k}{T}B_{z0}[J_{1}^{'}(Tr)-\frac{k_{z}}{k}\frac{J_{1}(Tr)}{Tr}]\sin(\phi)e^{-jk_{z}z}+c.c.,
\end{split}
\vspace{-10pt}
\end{align}

\noindent where $E_{z0}$ and $B_{z0}$ are the amplitude of longitudinal electric and magnetic fields respectively, $k=2\pi/\lambda_{0}$ is the wave number in vacuum and $\lambda_{0}$ is the wavelength of incident laser pulse, $k_{z}$ and $T$ are the longitudinal and transverse component of wave number inside a MPW, satisfying $k_{z}^{2}+T^{2}=k^{2}$. $J_{1}(x)$ is the bessel function of the first kind with integer order 1. By applying the same method to the plasma cladding region and using boundary condition that the tangential components of $E$ and $H$ should be continuous at $r=R_{0}$ ($R_{0}$ is the MPW radius). one obtains the eigenvalue equation \cite{s28}

\begin{figure}[!b]
\vspace{-10pt}
\includegraphics[width=7.5cm]{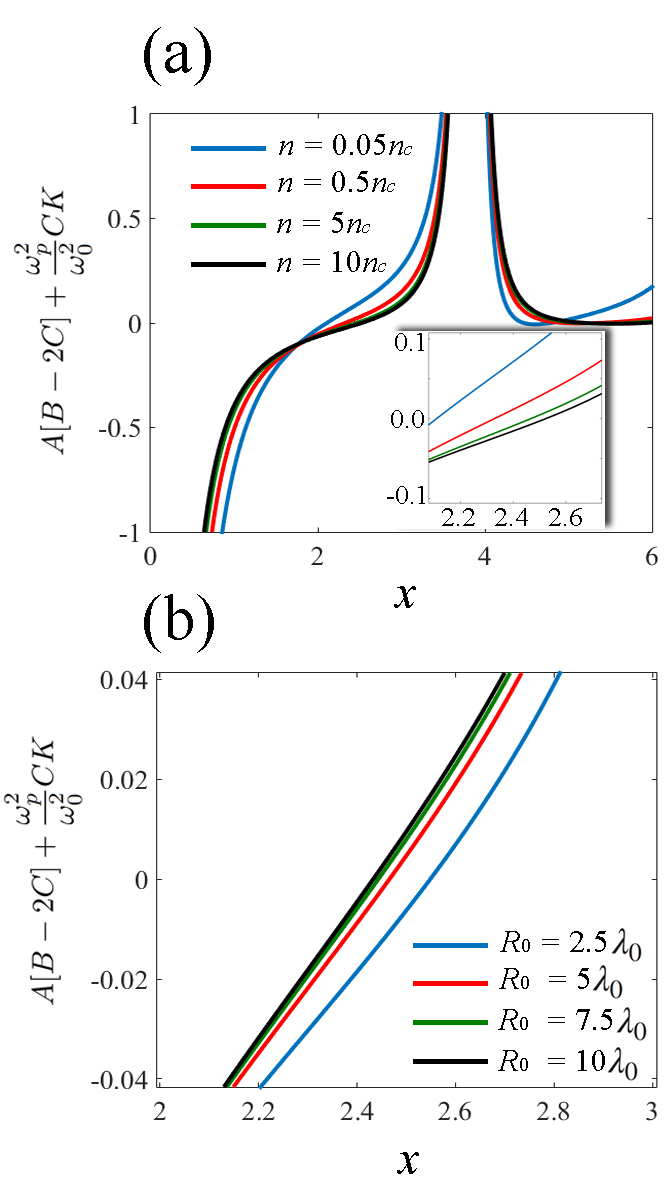}\caption{\label{f1} (Color online.) Left hand side of Eq.~(7) plotted against $x$ for different boundary plasma densities (a) and MPW radii (b), the inset in (a) presents the zoom-in of rectangular area inside the dashed deep red line. The MPW radius is fixed at $R_{0}=5\lambda_{0}$ in (a), and plasma cladding density is fixed at $n=5n_{c}$ in (b)}
\vspace{-20pt}
\end{figure}

\begin{align}
\vspace{-10pt}
\begin{split}
A[B-2C]+\frac{\omega_{p}^{2}}{\omega_{0}^{2}}CK=0,
\end{split}
\vspace{-10pt}
\end{align}

\noindent where
\begin{align}
\vspace{-10pt}
\begin{split}
A=J+K,
\end{split}
\vspace{-10pt}
\end{align}

\begin{align}
\vspace{-10pt}
\begin{split}
B=A+\frac{\omega_{p}^{2}}{\omega_{0}^{2}}[(1/y^{2})-K]
\end{split}
\vspace{-10pt}
\end{align}

\begin{align}
\vspace{-10pt}
\begin{split}
C=1/x^{2}+1/y^{2},
\end{split}
\vspace{-10pt}
\end{align}

\begin{align}
\vspace{-10pt}
\begin{split}
J=\frac{J_{2}(x)}{xJ_{1}(x)}, K=\frac{K_{2}(y)}{yK_{1}(y)}
\end{split}
\vspace{-10pt}
\end{align}

Here $y = \sqrt{4\pi^{2}R_{0}^{2}/\lambda_{p}^{2}-x^{2}}$, $\lambda_{p}$ is the plasma wavelength at the boundary, $K_{n}(y)$ is the modified Bessel function of the second kind with integer order $n$ ($n=1,2$). The first root of Eq.~(7) $x_{1}$ corresponding to the lowest optical modes, where transverse wave number $T=x_{1}/R_{0}$. It can be solved numerically as shown in Fig.~2.

Equations~(7-11) indicate once the laser frequency $\omega_{0}$ is settled, the roots of eigenvalue equation depend solely on MPW radius $R_{0}$ and plasma density on the boundary. However, as illustrated in Fig.~2, the value of $x_{1}$ varies little as long as the boundary plasma is overdense $n>n_{c}$, where $n_{c}=m_{e}\omega_{0}^{2}/4\pi e^{2}$ is the critical density, here $e$ and $m_{e}$ denotes the unit charge and electron mass, respectively. As a result, although the pre-pulse induced density expansion reduces the electron number that can be pulled out of channel wall, the electron dynamics in the MPW is only slightly modified and the underlying physics is almost the same. In addition, it should be noted we didn't consider the channel boundary density above $10 n_{c}$ in Fig.~2(a) because in real experiments, the laser can hardly penetrate into the area with $n>10 n_{c}$ due to finite density gradients.

By solving the eigenvalue equation Eqs.~(7-11), the phase velocity of the optical modes can be written as

\begin{align}
\vspace{-10pt}
\begin{split}
v_{p}=(1+\frac{x_{1}^{2}\lambda_{0}^{2}}{8\pi^{2}R_{0}^{2}})c,
\end{split}
\vspace{-10pt}
\end{align}

\noindent which relies solely on channel radius $R_{0}$. Here $c$ is the light velocity in vacuum.

\section*{III. ELECTRON ACCELERATION}

\subsection*{A. Acceleration by TM modes}

As the laser pulse propagating in the MPW, electrons in the plasma cladding are extracted into the channel due to wave breaking of Langmuir waves that are stimulated within the skin layer \cite{s31}. Those located in the right phase ($E_{z}<0$) can be accelerated by the TM modes. From Eqs.~(1-6), on can see the amplitude of longitudinal fields is smaller than transverse fields by a factor of $T/k\approx0.4\lambda_{0}/R_{0}$, so the energy coupled into longitudinal fields is negligible $T^{2}/k^{2}\ll1$. Therefore the amplitudes of the transverse field are roughly equal to the laser fields $E_{0}$ and $B_{0}$, and the peak acceleration gradient can be estimated as

\begin{align}
\vspace{-10pt}
\begin{split}
E_{m}\approx\frac{x_{1}a_{0}\lambda_{0}}{2\pi R_{0}}\frac{m_{e}c\omega_{0}}{e},
\end{split}
\vspace{-10pt}
\end{align}

\noindent where $a_{0}=eE_{0}/m_{e}c\omega_{0}$ is the normalized laser amplitude. The acceleration is limited by the dephasing effect owing to the superluminal phase velocity of the TM modes, which occurs when phase slippage equals to $\lambda_{0}/2$, resulting a dephasing distance

\begin{align}
\vspace{-10pt}
\begin{split}
L_{d}=\frac{4\pi^{2}R_{0}^{2}}{x_{1}^{2}\lambda_{0}},
\end{split}
\vspace{-10pt}
\end{align}

\noindent according to Eq.~(12). Therefore, the maximum electron energy can be estimated as

\begin{align}
\vspace{-10pt}
\begin{split}
\varepsilon_{m}\approx\frac{1}{2}eE_{m}L_{d}\approx\frac{2\pi^{2}R_{0}a_{0}}{x_{1}\lambda_{0}}m_{e}c^{2}.
\end{split}
\vspace{-10pt}
\end{align}

\begin{figure}[!b]
\vspace{-10pt}
\includegraphics[width=8.5cm]{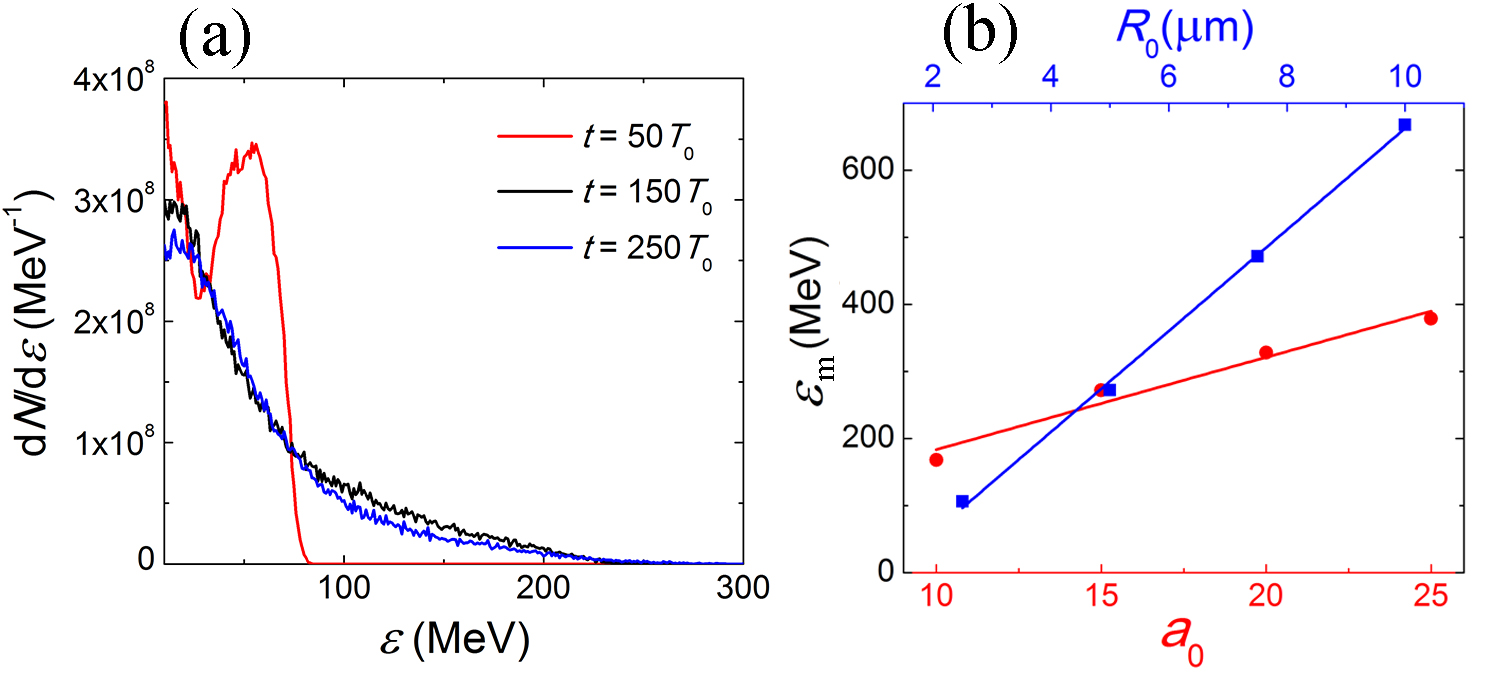}\caption{\label{f1} (Color online.) (a) Electron energy spectrum at different simulation times times for the $a_{0}=15$, and $R_{0}=5\mu$m, (b) the maximum electron energies plotted against $a_{0}$ (red) and $R_{0}$ (blue).}
\vspace{-20pt}
\end{figure}

In the following, we perform 3D PIC simulations using the code Virtual Laser Plasma Lab (VLPL) \cite{s32} to explore the acceleration process and test the scalings presented in Eq.~(15). The simulation box is sampled by grid steps $dx\times dy\times dz = 0.05\lambda_{0}\times 0.05\lambda_{0}\times 0.025\lambda_{0} $ in each direction, and the time step is $dt=0.02T_{0}$, where $\lambda_{0}=1\mu$m is the laser wavelength and $T_{0}=3.34$ fs is the laser cycle. A moving window is used to improve computational efficiency, with longitudinal dimension equals to $18\lambda_{0}$, the transverse dimensions of simulation box are chosen according to the MPW radius that employed in different runs. A Gaussian laser pulse is focused on the entrance of MPW, propagating towards $+z$ direction, with a temporal FWHM duration of 20 fs and a focus spot equals to $0.7R_{0}$. A hollow MPW with cladding-plasma density $n_{0}=5 n_{c}$ is employed in the simulations. The normalized laser amplitude $a_{0}$ in the range 10-25, and the inner radius of MPW $R_{0}$ in the range $2.5\mu$m to $10\mu$m are considered to examine the scaling laws obtained from theoretical model. In addition, a CP drive laser is considered in the main body of this paper, while a brief discussion on the effects of laser polarization can be found in Sec.~IV.D.

The typical electron energy spectrum that obtained in PIC simulation is shown in Fig.~3(a) for different simulation times with laser amplitude $a_{0}=15$ and inner radius $R_{0}=5\mu$m. According to Eq.~(14), the dephasing distance in this case is roughly 158 $\mu$m, which agrees well with our numerical results. Moreover, one can see a quasi-monoenergetic peak in the energy spectrum in the early stage ($L<158\mu$m), which gradually vanishes as the dephasing occurs, resulting a Maxwell-like distribution.

The maximum electron energies are plotted for different $a_{0}$ and $R_{0}$ in Fig.~3(b), which shows a linear dependence for both cases as predicted by Eq.~(15). Moreover, one can infer from Fig.~3(a) the total number of the electrons is on the order of $10^{10}$. Owing to the co-propagation dynamics and proper guiding of laser pulse by the MPW, such large amount of energetic electrons is promising to extract significant energy from the drive laser and produce abundant X-ray photons during the long interaction period.

\subsection*{B. Role of the ponderomotive force}

So far we have considered the electron acceleration by the longitudinal electric field of TM modes, however, the effects of nonlinear ponderomotive force also need to be addressed since the laser intensity that used to drive the X-ray source is normally in the highly relativistic regime ($a_{0}\gg1$). Under such circumstance, the ponderomotive force arises due to the spatial variation of intensity of the EM wave that in general pushes the charged particles out the regime of high intensity and exert pressure on the target \cite {s33}. The electron acceleration due to ponderomotive force has been broadly referred to as the "direct laser acceleration" \cite{s34,s35,s36,s37,s38} [in the following we use the term "ponderomotive laser acceleration (PLA)" to avoid possible confusing]. In our case, the electron acceleration by the laser pulses inside the MPW takes place in presence of both of TM modes and ponderomotive force, apparently it benefits from both mechanisms.

The major effect of PLA is altering the injection process, allowing the electrons to be injected within a much shorter distance. Since the phase velocity of TM modes are superluminal, electrons gradually falls behind the accelerating phase, the initial PLA greatly prolongs the dephasing distance, and therefore enhances the energies that acquired by the electrons. Take $a_{0}=15, R_{0}=10\lambda_{0}$ for example, in the absence of ponderomotive force, the electrons initially at rest would cause a relative phase slippage roughly equals to $0.13\lambda_{0}$ to gain relativistic energies ($\sim m_{e}c^{2}$) in order to be "trapped" by the EM wave. As a result, one would expect the dephasing distance is $444\mu$m (corresponding to a relative phase slippage equals to $0.37\lambda_{0}$) and maximum energy is 444 MeV. However, the PIC simulation results suggest otherwise as presented in Fig.~3(b). The maximum energy attained is roughly 1.5 times larger, which indicates the tremendous ponderomotive potential accelerates the electrons to relativistic energies within a much shorter distance that almost negligible compare with laser cycle.

However, the PLA becomes less and less important as the acceleration goes on. The underlying physics can be understood as follows, the energy attainted by electrons in unit time is $e(E_{0}v_{\bot}+E_{z}v_{z})$, where $v_{\bot}$ is the transverse electron velocity. The former term is the contribution from PLA, and the latter results from direct acceleration via TM modes. In the beginning, as the electrons are pulled out from the channel walls, they are accelerated via the laser electric fields to a transverse velocity that approaches light speed, i.e. $v_{\bot}\rightarrow c$, at this moment, the acceleration process is dominated by the PLA, since $E_{z}\approx T/kE_{0}\ll E_{0}$. However, after the injection ($v_{z}\rightarrow c$), the transverse velocity $v_{\bot}$ decreases rapidly due to an increasing of the gamma factor.

As illustrated by Fig.~4, for the electrons with energy above 60 MeV, the acceleration by TM modes becomes more important (i.e. $v_{\bot}/c<T/k$). It also suggests that the majority of energetic electrons we interested (in the area with darker color) possess small transverse velocities that the acceleration gradient of the TM modes is roughly 4 times larger than the maximum value can be provided by PLA. The latter, in addition, requires transverse electron motion stays anti-parallel to the laser electric field, which can not be always fulfilled by the majority of electrons. This is because the phase difference between $E_{r}$ and $E_{z}$ is $\pi/2$ as shown in Eqs.~(1) and (3), so that an electron with transverse velocity moving towards the axis are accelerated by PLA in the first half of acceleration bucket, while decelerated in the second half. As a consequence, we consider the acceleration procedure are dominated by the TM modes after the initial injection, though the contribution of PLA could explain small amount ($\sim1$pC) of high energetic electrons beyond the energy limit predicted by Eq.~(15) that observed in the simulations.

\begin{figure}[!t]
\vspace{-10pt}
\includegraphics[width=7.5cm]{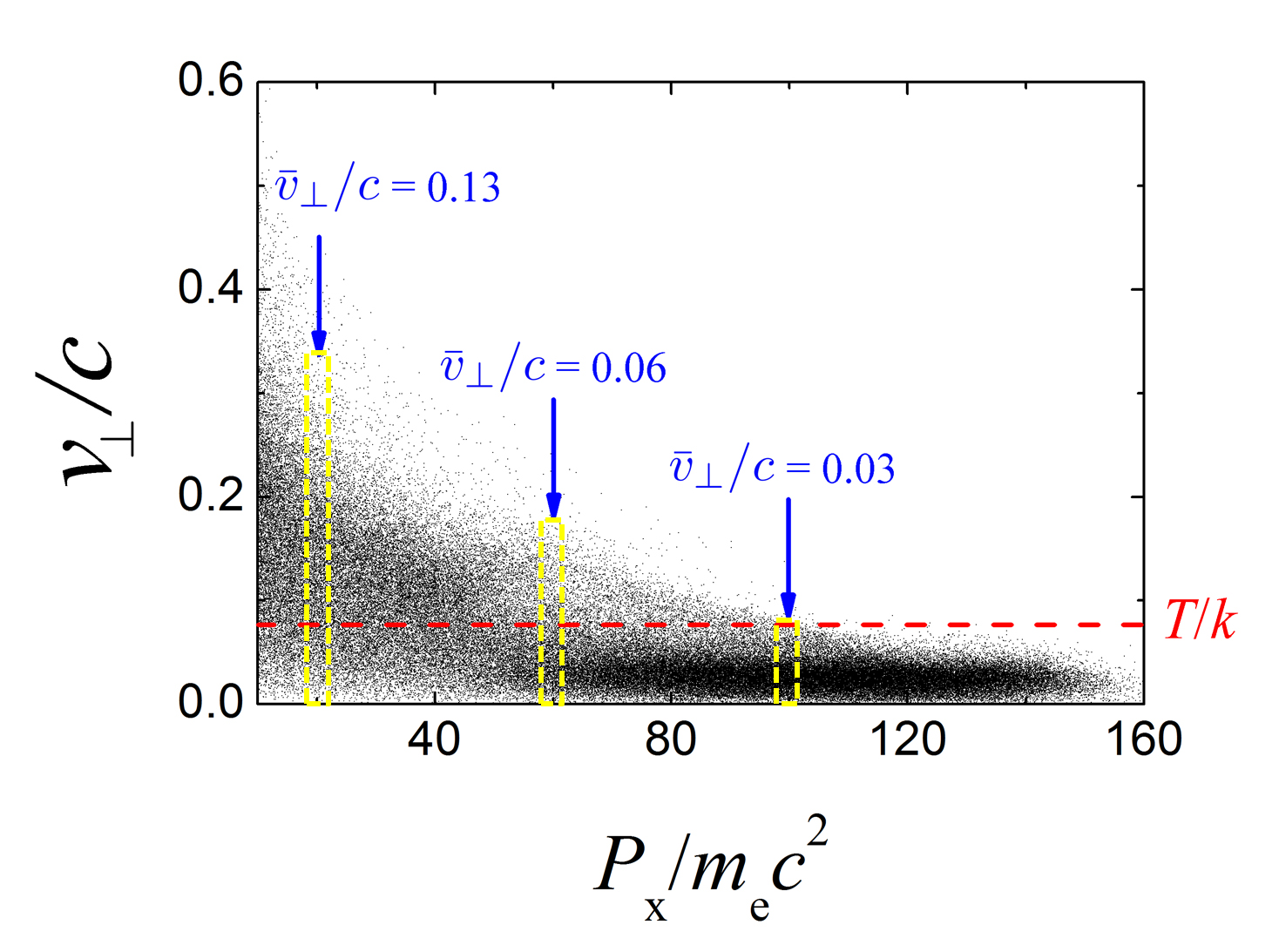}\caption{\label{f1} (Color online.) The electron transverse velocity dependence on longitudinal monmentum at propagation distance $L=50\mu$m, the simulation parameters are the same as in Fig.~3(a). The red dashed line presents the amplitude of longitudinal electric fields of TM modes normalized by the laser amplitude ($E_{z0}/E_{0}\approx T/k$), and average transverse velocity ($\overline{v}_{\bot}$) of electrons with longitudinal momentums around $P_{x}=$20, 60, and 100 MeV (electrons in the area marked by yellow rectangular) are shown.}
\vspace{-20pt}
\end{figure}

We emphasis that the (i) small radius ($<$ 10 microns) and (ii) relatively-long length ($\sim$ several hundreds of microns) are two important features which distinguish the MPW considered in this paper with many other studies \cite{s5,s12,s13,s25,a1,s39,s40} concerning electron acceleration in overdense plasma channels. The former (i) gives rise to significant longitudinal electric fields (by TM modes), so that the EM waves inside the MPW cannot be considered as plane wave any more; and the latter (ii) provides enough distance for the TM modes to dominate the electron acceleration process. In the remainder of this paper, we will discuss the dynamics of those energetic electrons, specifically the transverse wiggling due to the remarkable asymmetry of EM wave that induced by (i), and the generation of bright X-rays over long the interaction distance that enabled by (ii).

\section*{IV. X-RAY EMISSION}

\subsection*{A. Transverse electron dynamics}

It is well-known that a transverse wiggling force is essential to every synchrotron radiation source. However, it must be handled carefully to achieve a good performance of the X-ray source since the radiation divergence will be deteriorated if the force is too strong. In our case, the transverse force acting on the relativistic electrons can be calculated by Eqs.~(3-6)

\begin{align}
\vspace{-10pt}
\begin{split}
F_{\bot}\approx\frac{\sqrt{2}x_{1}^{2}\lambda_{0}^{2}}{8\pi^{2}R_{0}^{2}}eE_{0}J_{0}(Tr)ie^{-ik_{z}z}+c.c.,
\end{split}
\vspace{-10pt}
\end{align}

\noindent where $F_{\bot}=e\sqrt{(E_{r}-B_{\phi})^{2}+(E_{\phi}+B_{r})^{2}}$ is the force perpendicular to laser propagation direction, we have assumed the velocity of electrons close to light speed $\beta\rightarrow1$, and the factor $\sqrt{2}$ arises due to the CP laser employed. Apparently, unlike the plane EM waves, the optical modes in the MPW exert a remarkable net transverse force on the ultra-relativistic electrons, regardless the velocity. The small asymmetry between electric and magnetic fields, as suggested by Eq.~(16), can be adjusted by varying the MPW radius $R_{0}$, which provides an important degree of freedom to design the X-ray source.

In the presence of a wiggling force expressed by Eq.~(16), one would expect a near-axis electron [$J_{0}(Tr)\sim1$] undergoes a sinusoidal-like trajectory with the maximum transverse displacement equals to $r_{max}\approx4\sqrt{2}\pi R_{0}^{2}a_{0}/(x_{1}^{2}\lambda_{0}\gamma)$. So that the maximum opening angle of the radiation cone can be calculated as

\begin{align}
\vspace{-10pt}
\begin{split}
\Theta=\frac{2\pi r_{max}}{\lambda_{u}}\approx\frac{\sqrt{2}a_{0}}{\gamma},
\end{split}
\vspace{-10pt}
\end{align}

\noindent where

\begin{align}
\vspace{-10pt}
\begin{split}
\lambda_{u}\approx\frac{8\pi^{2}R_{0}^{2}}{x_{1}^{2}\lambda_{0}}
\end{split}
\vspace{-10pt}
\end{align}

\noindent is the wiggler spatial period that depended solely on the superluminal phase velocity $v_{p}$ of waveguide modes. From Eq.~(17), one can easily obtain the wiggling strength by

\begin{align}
\vspace{-10pt}
\begin{split}
K=\gamma\Theta\approx\sqrt{2}a_{0}.
\end{split}
\vspace{-10pt}
\end{align}

Equations~(18) and (19) describe the transverse electron dynamics in the context of synchrotron radiation, where the waveguide modes with superluminal $v_{p}$ is considered as a wiggler. Owing to the ultra-intense ($a_{0}\gg1$) laser pulses employed, the radiation contains many high harmonics, leading to a synchrotron-like broadband spectrum \cite{s41}. The critical frequency of the generated X-ray photons can be calculated as \cite{s33}

\begin{align}
\vspace{-10pt}
\begin{split}
\omega_{c}\approx\frac{3\sqrt{2}a_{0}\gamma^{2}x_{1}^{2}\lambda_{0}^{2}}{16\pi^{2}R_{0}^{2}}\omega_{0}.
\end{split}
\vspace{-10pt}
\end{align}

\subsection*{B. Properties of X-ray emissions}

Keeping in mind the possible applications \cite{s42,s43,s44,s45,s46} of hard x-ray photons, we focus on the angular distribution of radiation and its spectrum, along with the x-ray emission brillance. By plugging in the gamma factor in Eqs.~(17) and (20) with half of the maximum energy that expressed in Eq.~(15), one obtains the maximum opening angle and critical frequency as

\begin{align}
\vspace{-10pt}
\begin{split}
\Theta\approx\frac{\sqrt{2}x_{1}\lambda_{0}}{\pi^{2} R_{0}},
\end{split}
\vspace{-10pt}
\end{align}

\noindent and

\begin{align}
\vspace{-10pt}
\begin{split}
\omega_{c}\approx\frac{3\sqrt{2}\pi^{2}a_{0}^{3}}{16}\omega_{0}
\end{split}
\vspace{-10pt}
\end{align}

\noindent respectively. One can see that the divergence of X-ray emission scales as $R_{0}^{-1}$ and the photon energy scales as $a_{0}^{3}$. PIC simulations are conducted to test these scalings and provide more details on the X-ray emission properties, the results are shown in Figs.~5 and 6.

\begin{figure}[!b]
\vspace{-10pt}
\includegraphics[width=9cm]{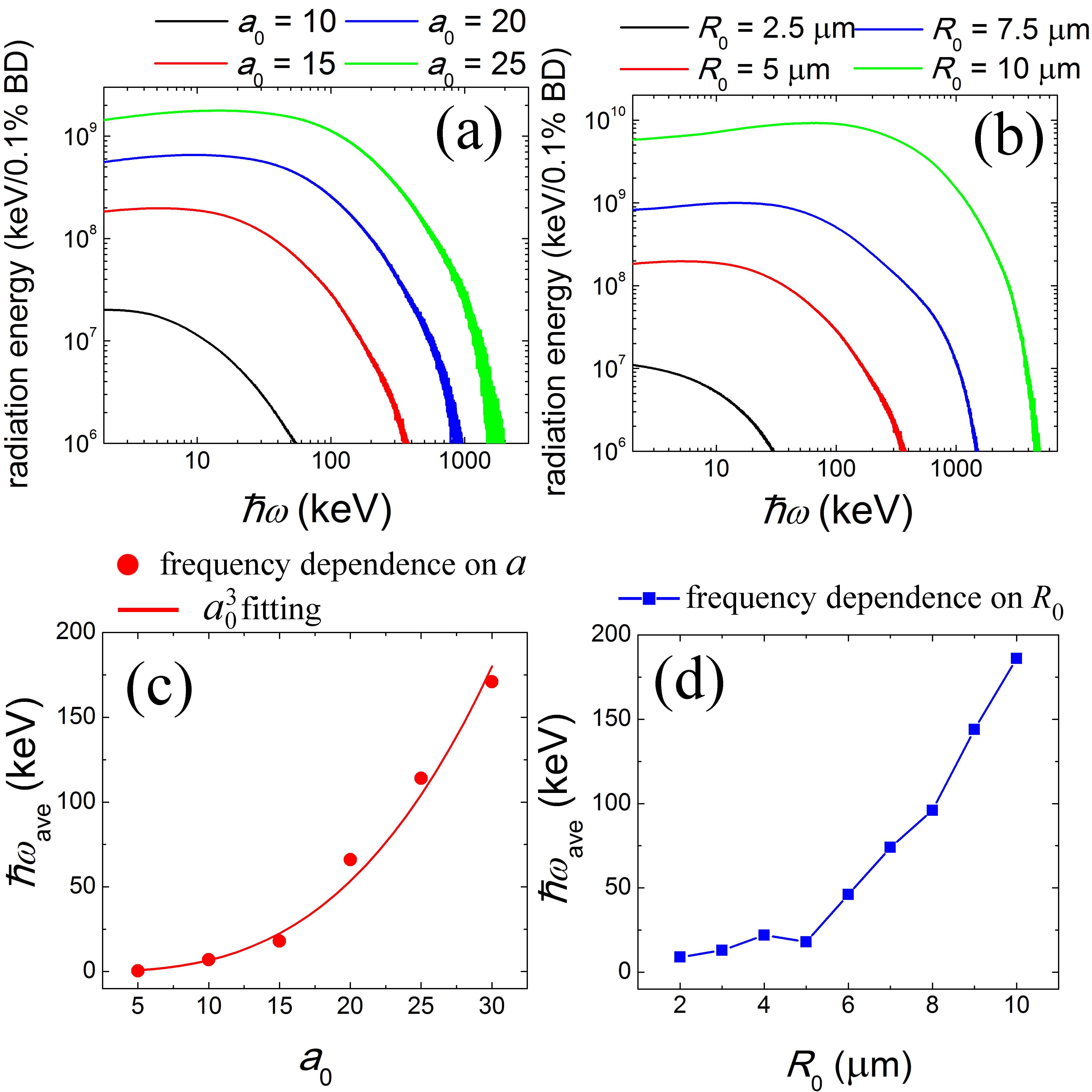}\caption{\label{f1} (Color online.) Radiation spectrum [radiation energy per $0.1\%$ bandwidth (BD) against photon energy)] for different laser amplitude $a_{0}$ (a) and MPW radii $R_{0}$ (b), as well as the average photon energy are plotted against $a_{0}$ (c) and $R_{0}$ (d). It should be noted when scanning different $a_{0}$, other parameters are remained the same (with $R_{0}=5\mu$m)], while for different $R_{0}$, we kept $a_{0}=15$ as a constant, but the spotsize of laser pulses ($\sim R_{0}$) and the length of MPW ($\sim R_{0}^{2}$) are changed accordingly.}
\vspace{-20pt}
\end{figure}

\begin{figure*}[]
\vspace{-10pt}
\includegraphics[width=16.0cm]{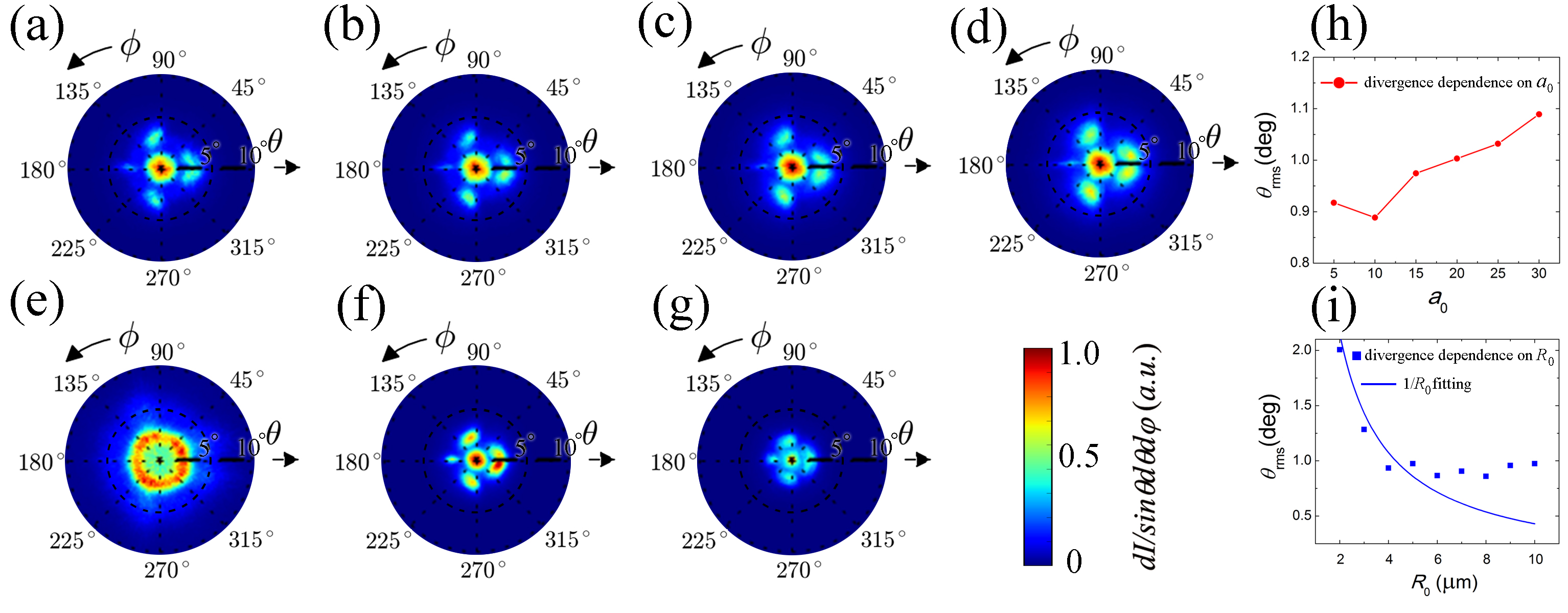}\caption{\label{f1} (Color online.)  Radiation angular distribution for different laser amplitude $a_{0}=$ 10(a), 15(b), 20(c), and 25(d) for fixed radius $R_{0}=5\mu$m as well as MPW radii $R_{0}=$ 2.5$\mu$m(e), 5$\mu$m(b), 7.5$\mu$m(f), and 10$\mu$m(g) for fixed laser amplitude $a_{0}=15$; and the r.m.s divergence of X-ray emission are plotted against $a_{0}$ (h) and $R_{0}$ (i). Here $\theta$ is the angle between photon emission direction and $z$ axis, and $\phi$ is the angle between photon emission direction and $x$ axis in the $x-y$ plane.}
\vspace{-10pt}
\end{figure*}

Figures~5(a-b) present the radiation spectrum of the X-ray photons for different laser plasma parameters. For every case presented in Fig.~5, typical synchrotron-like broadband spectrums are observed, with the photon energies covering a range that from hundreds of keVs to MeVs, depending on the laser plasma parameters used. The "long tails" on the radiation spectrum are mainly produced due to the Maxwell-like distribution of the electron energy in MPW, as shown in Fig.~3.

The average photon energy (total radiation energy divided by photon number) are plotted against $a_{0}$ and $R_{0}$ in Fig.~5(c-d). The dependence of radiation frequency on laser amplitude can be fitted with a cubic scaling as predicted by Eq.~(22). On the other hand, one can see that when the channel radius is small, the radiation frequency seems independent with $R_{0}$ as expected, but the scaling changes when $R_{0}$ exceeds $5\lambda_{0}$, the radiation frequency increases almost linearly with channel radius. This indicates the excitation of higher order waveguide modes, (which will be qualitatively discussed in the next subsection) so that the transverse force cannot be estimated by considering the lowest modes alone.

Similarly, in Fig.~6 we presents the angular distributions of X-ray emission intensity for different laser plasma parameters. Figs.~6(a-d) illustrate comparisons for different laser amplitude (for a fixed MPW radius $R_{0}=5\mu$m), and (b),(e-g) for different MPW radii (for a fixed laser amplitude $a_{0}=15$). In general, well-collimated X-ray emission with root mean square (RMS) opening angles $\theta_{rms}$ below 2 degrees are obtained for all the cases, which can be attribute to the relatively-small transverse wiggling force [from $10^{-3}eE_{0}$ to $0.02eE_{0}$ for the parameters considered in this paper, according to Eq.~(16)]. In addition, Fig.~6(i) indicates that the divergence of radiation is independent with $a_{0}$ as suggested by Eq.~(21), but again, the numerical observations depart from the $1/R_{0}$ scaling when the channel grows bigger ($R_{0}> 5\lambda_{0}$) as shown by Fig.~6(h).

Finally, Fig.~7 shows sans of photon yield ($N_{p}$) and overall (laser-to-photon) efficiency ($\eta$) plotted over laser amplitude $a_{0}$ and MPW radii $R_{0}$. Typically, $10^{10}\sim10^{11}$ photons are produced in a single shot by the presented X-ray source. Considering the duration of X-ray emission governed by the laser duration, and a spot size roughly equals to the cross-section area of MPW, the peak brilliance reaches $\sim10^{23}$ photons/s/mm$^{2}$/mrad$^{2}$/0.1$\%$bandwidth. The overall efficiency is on the order of $10^{-5}$, which increases rapidly with both $a_{0}$ and $R_{0}$.

Figure~7(a-b) indicates the photon yield and efficiency can be roughly fitted by $a_{0}^{2}$ and $a_{0}^{3}$, respectively. This can be understand as follows: assuming the energy ratio transformed form laser to energetic electrons is the same, so the electron number produced scales as $N_{e}\propto a_{0}$ [because average electron energy $\overline{\varepsilon}\propto a_{0}$ as indicated in Eq.~(15)], while radiation power of a single electron scales as $P_{rad}\propto \gamma^{2}F_{\bot}^{2}\propto a_{0}^{4}$, so that the photon yield scales as $N_{p}\propto N_{e}P_{rad}/\hbar\omega_{ave}\propto a_{0}^{2}$ and the overall efficiency $\eta\propto N_{e}P_{rad}/a_{0}^{2}\propto a_{0}^{3}$.

On the other hand, the relation between $N_{p}$, $\eta$ and MPW radii $R_{0}$ presented in Fig.~7(c-d) is much more difficult to quantitatively analyze. However, the simulations suggest a very favorable scalings with $R_{0}$, the photon yield and efficiency are both rapidly increasing as the channel turns bigger. We believe besides the increasing in the gamma factor of electrons, the excitation of high order modes also plays an important role that enhances the transverse wiggling force, and therefore the X-ray emission.

\begin{figure}[!t]
\vspace{-10pt}
\includegraphics[width=8.5cm]{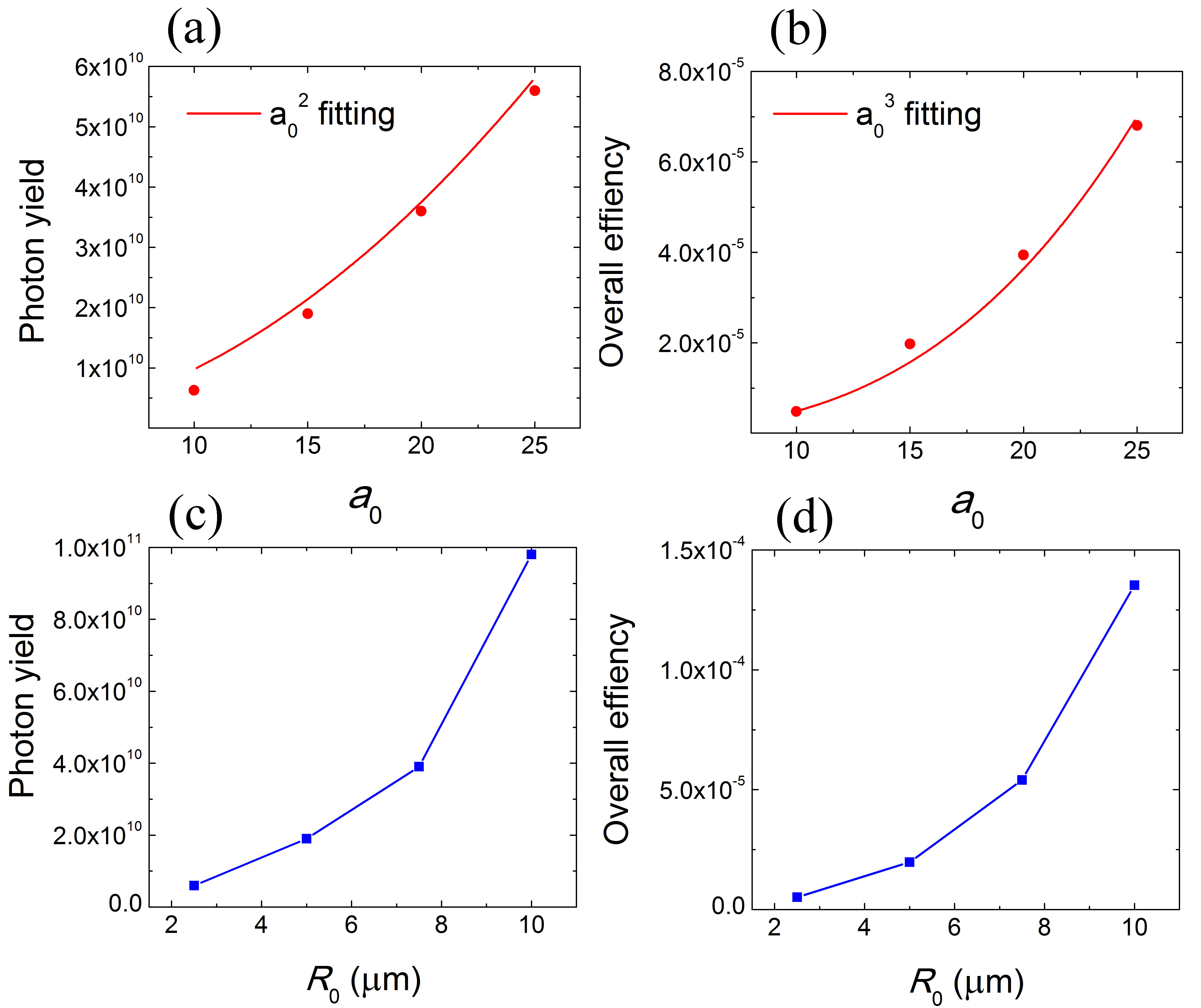}\caption{\label{f1} (Color online.) The dependence of photon yield (a),(c) and overall efficiency (b),(d) on laser amplitude $a_{0}$ (a-b) for fixed MPW radius $R_{0}=5\mu$m and MPW radii $R_{0}$ (c-d) for fixed laser amplitude $a_{0}=15$. the red curve in (a) and (b) presents fitting of $a_{0}^2$ and $a_{0}^3$, respectively}
\vspace{-20pt}
\end{figure}

\subsection*{C. Role of high order modes}

In the theoretic model described in Sec.~II, we only consider the lowest waveguide modes in the MPW since the majority of the incident laser energy is coupled into these modes. The analytic scalings obtained above are adequate to describe the observed phenomenon in most cases. However, it fails to explain the behaviors of radiation behaviors in the big channels ($R_{0}>5\lambda_{0}$) as shown in Fig.~5-7, which brings the problem of multimode effect.

Normally, as a laser pulse propagates in the MPW, many modes are excited simultaneously, and the electron motion is subjected to the combined fields of these modes. The index (n) of waveguide modes is corresponding to the n-th root ($x_{n}$) of eigenvalue equation Eq.~(7), which increases with n. Therefore, the higher order modes possess larger transverse wave number ($T_{n}=x_{n}/R_{0}$), leading to a higher phase velocity, so that the electrons are wiggled with a higher frequency. Also, since more energy is coupled into the longitudinal fields ($E_{z}\&B_{z}\propto T_{n}/k$), the transverse asymmetry in EM wave is larger, resulting a much stronger transverse wiggling force. As a result, when higher order modes are excited in MPW, the radiation power is enhanced, and the photon energy increases, while the divergence deteriorates.

The energy proportion coupled into different modes is mainly dependant on the MPW radius, a more detailed investigation can be rely on PIC simulations, as illustrated by Fig.~8, which show the longitudinal electric fields $E_{z}$ at the same cross section for different values of $R_{0}$. We know that different modes correspond to different spatial distributions that governed by Bessel functions. Figure~8 indicates the multimode effect becomes important when $R_{0}$ exceed $5\lambda_{0}$, which coincides with the behaviors of X-ray photon emission that presented in Figs.~5$\sim$7. As a consequence, the excitation of high order mode provides us a convenient degree of freedom ($R_{0}$) to adjust the X-ray photon spectrum and radiation power.

\begin{figure}[!t]
\vspace{-10pt}
\includegraphics[width=8.5cm]{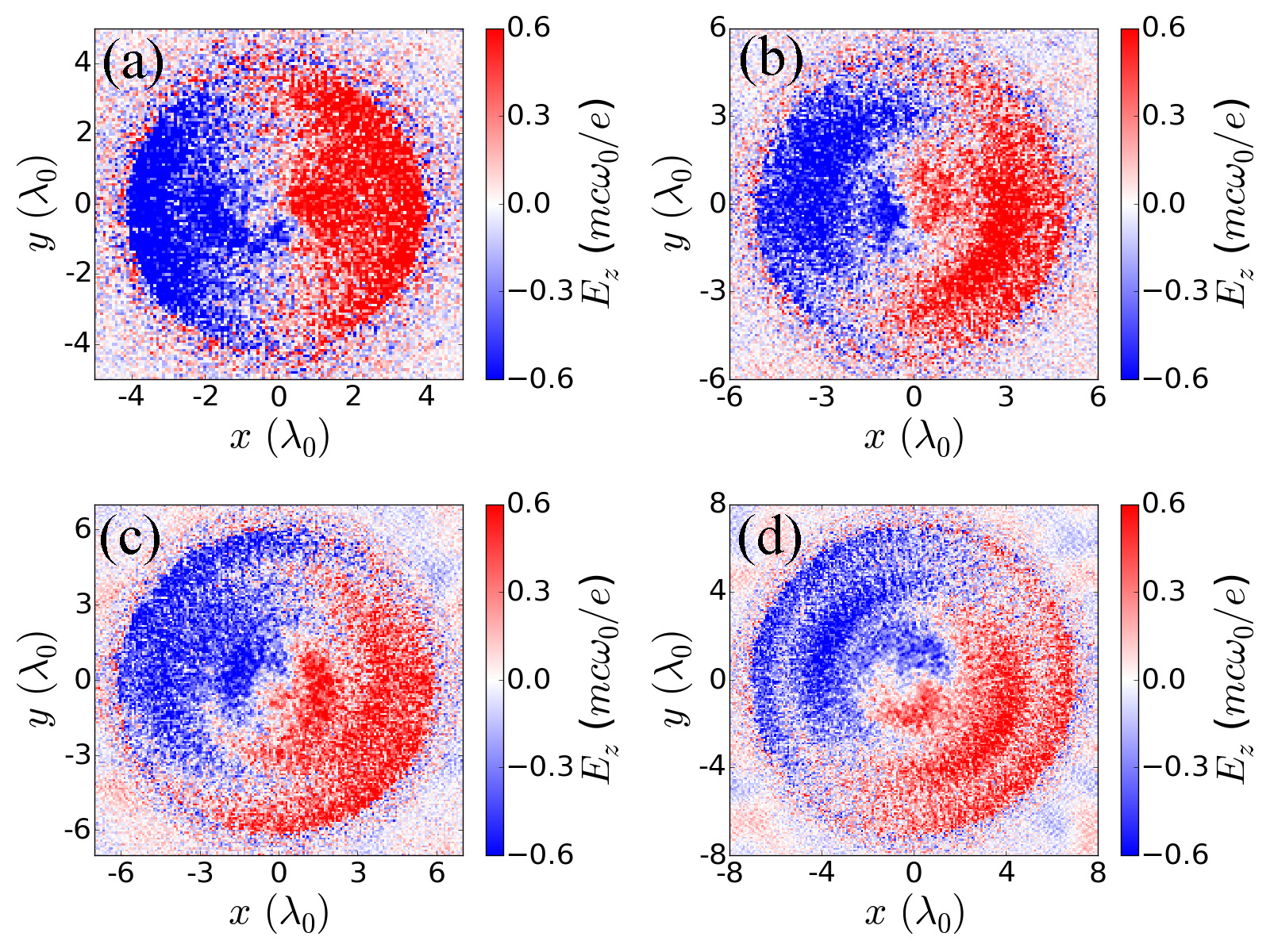}\caption{\label{f1} (Color online.) The Longidinal electric field distribution at cross-section $L=40\mu$m at simulation time $t=40T_{0}$ for different MPW radii $R_{0}=4\lambda_{0}$ (a), $5\lambda_{0}$ (b), $6\lambda_{0}$ (c), $7\lambda_{0}$ (d).}
\vspace{-20pt}
\end{figure}

\subsection*{D. Role of laser polarization}

When a linearly polarized (LP) laser pulse is used to drive the considered MPW X-ray source, electrons form a train of short bunches on each side of MPW boundaries in the polarization direction instead of a helical beam. The acceleration and wiggling processes are almost the same. However, the main difference lies in the angular distribution of X-ray emission as presented in Fig.~9(a). Most of the produced X-ray photons are emitted in the laser polarization direction $x$ ($\phi=0$), concentrated in 2 peaks around 3 degrees at each side of the propagating axis. In addition, the X-ray photons has a lower average energy in the LP case as shown in Fig.~9(b), and by integrating the spectrum, one can see the total photon number is decreased by a factor of three.

\begin{figure}[!t]
\vspace{-10pt}
\includegraphics[width=7.5cm]{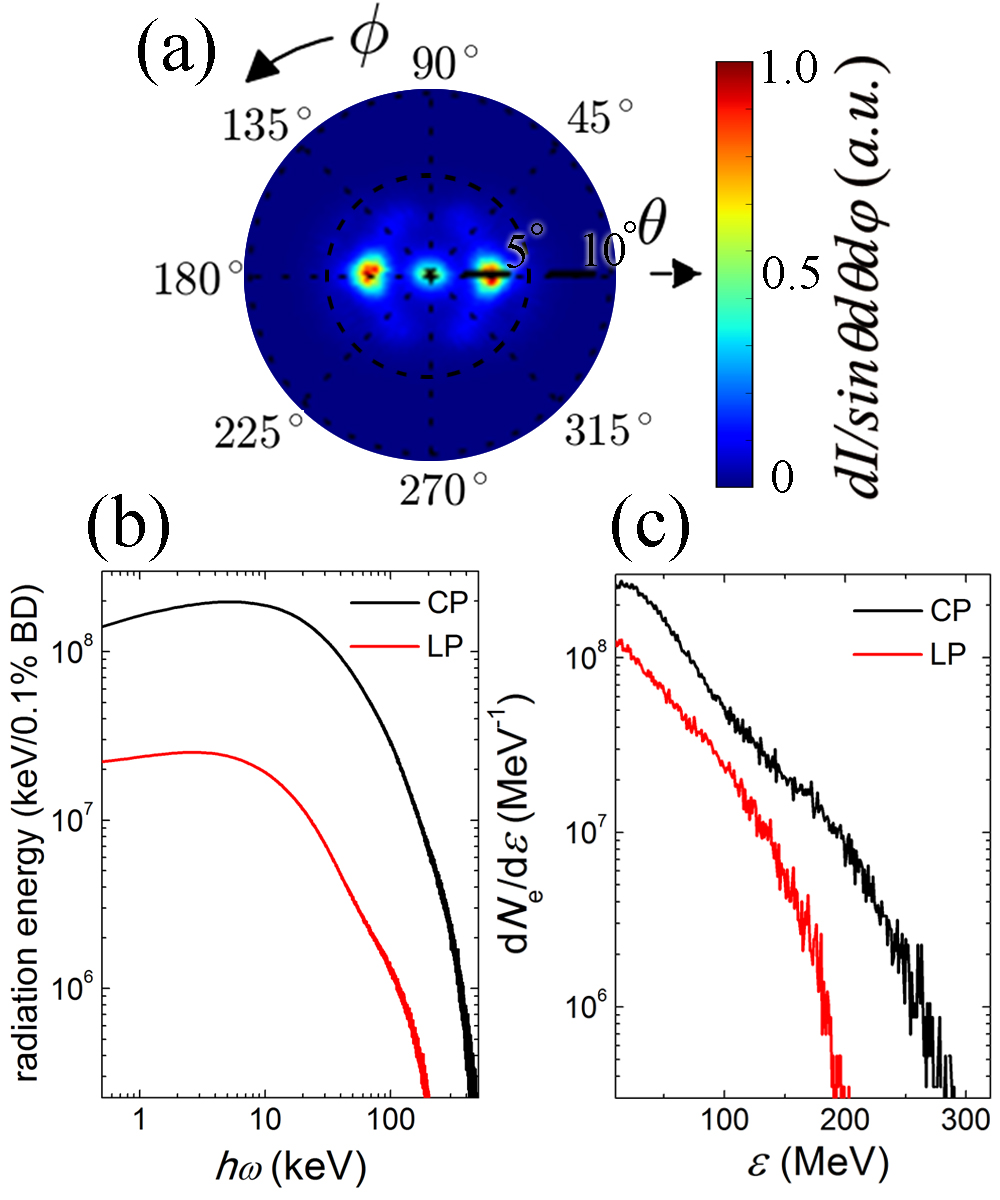}\caption{\label{f1} (Color online.) (a) The angular distibution of X-rays from a MPW illuminated by a LP laser with other parameters the same as in Fig.~6(b), and (b) a comparison of photon spectrum for CP (black line) and LP (red line) lasers against the propagating distance, the inset of (b) shows the electron spectrum in both cases.}
\vspace{-20pt}
\end{figure}

The above effects can be understood as a consequence of the reduction in both electron number and maximum energy, as illustrated in Fig.~9(c). Since the LP laser only contains $E_{x}$ component, the acceleration bucket is much smaller, and the accelerating gradient is slightly lower owing to lacking of the contribution from $E_{y}$ component. When a CP laser propagating in the MPW, the accelerating field $E_{z}$ contains two degenerated TM modes, with the same phase velocity but separated by a phase of $\pi/2$, i.e.

\begin{align}
\vspace{-10pt}
\begin{split}
E_{z}=E_{z0}J_{1}(Tr)\sin(\phi)[e^{-j(k_{z}z-\omega t)}+e^{-j(k_{z}z-\omega t+\frac{\pi}{2})}]+c.c..
\end{split}
\vspace{-10pt}
\end{align}

\noindent The composition is a helical accelerating structure, while for LP laser, only one component in the square bracket of Eq.~(23) exists, so the average acceleration gradient is $\sqrt{2}$ times smaller.

\section*{V. CONCLUSIONS AND DISCUSSIONS}

To summarize, using 3D PIC simulations, we studied the interaction of electrons with EM wave in a MPW in the ultra-relativistic regime, focusing on the synchrotron X-ray radiation that emitted by the self-generated energetic electron bunches wiggled by the waveguide modes. The resulted X-ray emission has promising features including high brightness, broad bandwidth, and low divergence. The underlying physical precesses are addressed which can be attributed to the fierce acceleration via the longitudinal electric field of TM modes, as well as the slight asymmetric structure of transverse optical mode components. The hard X-ray emission in the energy range of hundreds of keVs could reach a brilliance that on the order magnitude of $10^{23}$ photons/s/mm$^{2}$/mrad$^{2}$/0.1$\%$bandwidth. The typical RMS opening angle is below 2 degrees, and total photon yield is $10^{10}\sim10^{11}$ depending on the parameters employed, resulting a laser-to-photon efficiency roughly $10^{-5}$.

In addition, the X-ray properties for different laser and MPW parameters are investigated, specifically on the laser amplitude $a_{0}$ and MPW radius $R_{0}$. The common trend is that as $a_{0}$ increases, the radiation frequency and overall efficiency increase with $a_{0}^{3}$, photon yield scales as $a_{0}^{2}$, and the divergence remains unchanged as predicted by the analysis model.

On the other hand, the dependence of X-ray properties on the MPW radius is more complex. As long as $R_{0}$ is small enough ($R_{0}<5\lambda_{0}$), the EM wave in MPW can be treated by only considering the lowest waveguide modes. Under such circumstance, the radiation frequency is independent with $R_{0}$ and divergence scales as $1/R_{0}$ as predicted by the theoretic model. However, when $R_{0}$ exceeds some critical value around $5\lambda_{0}$, the multimode effect comes into play. In this case, as $R_{0}$ increases, the radiation power, photon energy as well as the efficiency are enhanced, while the RMS opening angle depart from the $1/R_{0}$ scaling and saturates at around 1 degree.

Finally, the effects of the laser polarization are addressed. As a LP laser interacting with the MPW, the radiation pattern is modified so that the photons are mainly emitted in polarization direction. In the mean time, the total photon yield and radiation frequency is decreased compared with the case with a CP driving laser of the same normalized amplitude.

\begin{acknowledgments}
This work is supported by DFG Transregio TR18, EU FP7 EUCARD-2 projects and National Natural Science Foundation of China (No.11505262, No.11125526, and No.11335013).
\end{acknowledgments}

\end{document}